\documentclass[prb,aps, twocolumn,amsmath,amssymb,amsfonts, superscriptaddress]{revtex4}
\usepackage[german,american,english]{babel}
\usepackage{graphicx}
\usepackage{graphics}
\usepackage{dcolumn}
\usepackage{bm}
\usepackage{amssymb}
\usepackage{amsmath}
\usepackage{amsfonts}
\usepackage{epsfig}

\newcommand {\bkt} [1] {\langle #1 \rangle}

\newcommand {\pd} [2] {\frac{\partial #1}{\partial #2}}
\newcommand {\td} [2] {\frac{d #1}{d #2}}

 \newcommand {\beq}{\begin{equation}}
\newcommand {\eeq}{\end{equation}}
\begin{document}
\title{Linear response theory of interacting topological insulators}
\author{Dimitrie Culcer}
\affiliation{ICQD, Hefei National Laboratory for Physical Sciences at the Microscale, University of Science and Technology of China, Hefei 230026, Anhui, China} 
\affiliation{Kavli Institute for Theoretical Physics, University of California, Santa Barbara, CA 93106-4030}

\begin{abstract}
Chiral surface states in topological insulators are robust against interactions, non-magnetic disorder and localization, yet topology does not yield protection in transport. This work presents a theory of interacting topological insulators in an external electric field, starting from the quantum Liouville equation for the many-body density matrix. Out of equilibrium, topological insulators acquire a current-induced spin polarization. Electron-electron interactions renormalize the non-equilibrium spin polarization and charge conductivity, and disorder in turn enhances this renormalization by a factor of two. Topological insulator phenomenology remains intact in the presence of interactions out of equilibrium, and an exact correspondence exists between the mathematical frameworks necessary for the understanding of the interacting and non-interacting problems.
\end{abstract}
\date{\today}
\maketitle

\section{Introduction}

The understanding of insulating behavior has been revolutionized by the landmark discovery of topological insulators (TI), \cite{KaneMele_QSHE_PRL05, Hasan_TI_RMP10, Qi_TI_RMP_10, Moore_TRI_TI_Invariants_PRB07} which are bulk band insulators with spin-orbit induced conducting states on the surface (3D) or edge (2D). These states are a manifestation of $Z_2$ \textit{topological} order: topology guarantees the existence in equilibrium of a crossing of bands connecting time-reversal invariant momenta, which is robust against smooth time-reversal invariant perturbations such as non-magnetic disorder and electron-electron interactions. The surface states of 3D TI are described by a Rashba Hamiltonian \cite{BychkovRashba_JETP84} with Dirac-cone like dispersion, and are gapless and chiral, with a well-defined spin texture (spin-momentum locking.) They carry a $\pi$ Berry phase, which protects against back-scattering and thus localization, and is associated with Klein tunneling, a half-quantized anomalous Hall effect \cite{ZangNagaosa_TI_Monopole_PRB10} and a giant Kerr effect. \cite{Tse_TI_GiantMOKE_PRL10} Nontrivial topology makes TI a platform for the observation of Majorana fermions \cite{Fu_Proximity_Majorana_PRL08} and for topological quantum computing. \cite{SDS_TQC_RMP08}

The rise of topological insulators is following a close parallel to the rise of graphene a short time ago. Three-dimensional topological insulators have grown from non-existence to a vastly developed mature field involving hundreds of researchers practically overnight. Within this time span, chiral surface states started out as a mere theoretical concept, were predicted to exist in several materials and were subsequently imaged.\cite{Hasan_TI_RMP10, Qi_TI_RMP_10} Unlike graphene, the Hamiltonian of topological insulators is a function of the real spin, rather than a sublattice pseudospin degree of freedom. This implies that spin dynamics is qualitatively different from graphene. Moreover, the twofold valley degeneracy of graphene is not present in topological insulators. Despite the apparent similarities, the study of topological insulators is thus not a simple matter of translating results known from graphene. Due to the dominant spin-orbit interaction, topological insulators are also qualitatively different from ordinary two-dimensional spin-orbit coupled semiconductors.

The topological order present in TI is a result of one-particle physics. In light of this, we recall that electron-electron interactions modify the effective mass, heat capacity, and ground state energy of solids, as well as the response of solids to external magnetic fields. \cite{Doniach} In fact, electron-electron interactions can lead to spontaneous magnetism in itinerant electron systems. The best-known example of this effect is Pauli paramagnetism in interacting electron systems. It is known from Fermi liquid theory that the Pauli paramagnetic susceptibility is enhanced by electron-electron interactions. This can be derived rigorously using various types of linear-response formalisms, such as diagrammatic Kubo linear response theory, the Keldysh kinetic equation formalism, or density-matrix formalisms based on the Liouville equation. Interaction effects in systems with \textit{strong spin-orbit interactions} have been studied in 2D TI \cite{Levin_FracTI_PRL09, Wu_QSH_Interact_11} and 3D TI, \cite{Varney_TI_ee_QPT_PRB10, Seradjeh_TI_XtnCond_PRL09, Behnia_Bi_Frac_Science07, Seradjeh_Bi_HiB_PRL09, Sun_TI_SponSymBrk_PRL09, Ran_TI_Stripes_10, Raghu_HlcLqd_PRL10} and previously in spin-orbit coupled semiconductors. \cite{ChenRaikh_PRB99, Glazov_NATO02, Shekhter_2DEG_ESR_ee_PRB05, Hankiewicz_PRB06, Tse_SO_Drag_PRB07, Zak_2DEG_SpinSusc_ee_PRB10} In topological insulators the focus has been on phenomena in equilibrium and in the quantum Hall regime. \footnote{Remarkably, Ostrovsky \textit{et al.}, Phys.\ Rev.\ Lett. \textbf{105}, 036803 (2010), showed that interactions can fully localize surface states in \textit{strongly disordered} TI.} 

In the mean time, transport in topological insulators has made enormous strides recently. \cite{Culcer_TI_PhysE11} In initial experimental efforts, it appeared impossible to identify any signatures whatsoever of the elusive surface states. Yet lately experimental work on transport in topological insulators has begun to advance at a brisk pace, and is without doubt entering its heyday, in the way ARPES and STM work did two years ago. A beautiful experiment \cite{Analytis_Bi2Se3_QL_SfcStt_NP10} recently detected the topological surface states of Bi$_2$Se$_3$, in which Sb was partially substituted for Bi to reduce the bulk carrier density to $10^{16}$cm$^{-3}$. At large magnetic fields the surface states were clearly seen, with Shubnikov-deHaas oscillations depending only on the perpendicular magnetic field, and oscillatory features growing with increasing field. Another work showed that carrier densities can be tuned over a wide range with a back gate. \cite{Chen_Bi2Se3_Tunable_PRB11} A more recent experimental breakthrough \cite{Sacepe_TI_NS_11} investigated surface transport in thin films of Bi$_2$Se$_3$ of thickness $\approx$ 10nm, observing Landau levels that evolve continuously from electron-like to hole-like. In another breakthrough, Kim \textit{et al.} \cite{Kim_TI_Gate_MinCond_11} studied Bi$_2$Se$_3$ surfaces in samples with thicknesses of $<10$nm, using a gate electrode to remove bulk carriers entirely and take both surfaces through the Dirac point simultaneously. Ambipolar transport was observed with with well-defined $p$ and $n$ regions, together with a minimum conductivity of the order of $e^2/h$, reflecting the presence of electron and hole puddles. Exciting developments in HgTe transport have also been reported. \cite{Buettner_HgTe_ZeroGapQW_NP11, Bruene_StrainedHgTe_QHE_PRL11}

Due to spin-momentum locking, the charge current flowing on the surface of a TI is intimately linked to its spin polarization. \cite{Culcer_TI_Kineq_PRB10} Firstly, it is evident that an \textit{out-of plane} spin polarization can be generated by a magnetic field or magnetization. However, an \textit{in-plane} magnetic field cannot generate an in-plane spin polarization for a Dirac cone: it simply shifts the origin of the cone and can be removed by a gauge transformation.\cite{Zyuzin_TI_B_parallel_QPT_PRB11} On the other hand, the combination of spin-momentum locking plus an electric field can be understood as a net effective magnetic field, which is in the plane of the TI, and generates an in-plane spin polarization. 

This paper presents a study of the role of electron-electron interactions in topological insulators in an electric field, and their effect on the spin polarization generated electrically \textit{in} the plane of the TI. A fundamental question is whether basic TI phenomenology survives interactions \textit{out of equilibrium}. It is known that in transport topology only protects against \textit{back}-scattering. Topological protection stems from time reversal symmetry, whereas transport is inherently irreversible. Therefore robustness against electron-electron interactions in equilibrium does not translate into the same robustness in transport. I will demonstrate that the effect of interactions can be absorbed by a renormalization of the non-interacting charge conductivity and spin polarization, and the response is qualitatively the same. Topological insulator phenomenology therefore remains unchanged by electron-electron interactions in the steady state.


A multiband matrix formulation is imperative to capture interband dynamics and disorder effects, which give a nontrivial factor of two to the renormalization factor appearing in the charge current and spin polarization. This paper uses an alternative matrix formulation of linear response theory, which contains the same physics as conventional approaches and is potentially more transparent, relying on the quantum Liouville equation in order to derive a kinetic equation for the density matrix. This theory was first discussed for graphene monolayers \cite{Culcer_Gfn_Transp_PRB08} and bilayers, \cite{Culcer_Bil_PRB09} and was recently  extended to topological insulators including the full scattering term to linear order in the impurity density. \cite{Culcer_TI_Kineq_PRB10} Peculiarities of topological insulators, such as the absence of backscattering, which reflects the $\pi$ Berry phase and leads to Klein tunneling, are built into this theory in a transparent fashion. In this work, the formalism of Ref.\ \onlinecite{Culcer_TI_Kineq_PRB10} is extended to account for electron-electron interactions via a mean-field approach. Since transport in non-interacting systems was studied in that work, only minimal overlaps required for consistency have been retained in this article. It is assumed that $T = 0$ so that electron-electron \textit{scattering} is absent. The theory assumes $\varepsilon_F \tau_p/\hbar \gg 1$, where $\varepsilon_F$ is the Fermi energy, located in the surface conduction band, and $\tau$ the momentum relaxation time. The physics considered here is distinct from spin-Coulomb drag, \cite{Amico_PRB00, Tse_SO_Drag_PRB07} which requires electron-electron \textit{scattering}, and from previous work on transport in \textit{non-interacting} TI. \cite{Culcer_TI_Kineq_PRB10} 

Electron-electron interaction effects have also been studied in graphene transport. \cite{SDS_Gfn_RMP11} The interaction physics discussed here is to be distinguished from that of graphene, since, as stated above, graphene is a multivalley system, and its Hamiltonian is a function of the pseudospin, due to the sublattice degree of freedom, rather than the real spin. It is also important to realize that the mean-field Hartree-Fock treatment of interactions is particularly advantageous in topological insulators, because formulating a large-$N$ renormalization group expansion is  a challenging task. This is because, whereas in graphene the spin and valley degeneracies yield $N =g_s g_v = 4$, but in topological insulators $N = g_s g_v = 1$. 

The outline of this paper is as follows. In Sec.~\ref{sec:Heff} a general effective Hamiltonian for interacting systems is introduced. The dynamics of the density matrix in interacting systems are discussed in a mean-field formulation in Sec.~\ref{sec:DM}. Following that, the effective one-particle kinetic equation is derived in Sec.~\ref{sec:KinEq}. This is then solved so as to obtain the correction to the conductivity and its enhancement due to disorder, followed by a brief discussion of current TIs, a summary and conclusions.

\section{Effective Hamiltonian for Interacting Systems}
\label{sec:Heff}

The many-body Hamiltonian is
\begin{equation}
\arraycolsep 0.3 ex
\begin{array}{rl}
\displaystyle H = & \displaystyle \sum_{\alpha\beta} H_{\alpha\beta} c^\dag_\alpha c_\beta + \frac{1}{2} \sum_{\alpha\beta\gamma\delta} V^{ee}_{\alpha\beta\gamma\delta} c^\dag_\alpha c^\dag_\beta c_\gamma c_\delta \\ [3ex]
= & \displaystyle H^{1e} + V^{ee}.
\end{array}
\end{equation}
The two-particle matrix element $V^{ee}_{\alpha\beta\gamma\delta}$ in a general basis $\{ \phi_\alpha({\bm r}) \}$ is given by
\begin{equation}
V^{ee}_{\alpha\beta\gamma\delta} = \int d^3r \int d^3r' \, \phi^*_\alpha({\bm r})\phi^*_\beta({\bm r}') V^{ee}({\bm r} - {\bm r}') \phi_\delta({\bm r})\phi_\gamma({\bm r}').
\end{equation}
Hermiticity implies $V_{\alpha\beta\gamma\delta} = V^*_{\gamma\delta\alpha\beta}$ and identity of electrons $V_{\alpha\beta\gamma\delta} = V_{\beta\alpha\delta\gamma}$. The antisymmetrized form is
\begin{equation}
\tilde{V}_{\alpha\beta\gamma\delta} = \frac{1}{2} \, (V_{\alpha\beta\gamma\delta} - V_{\alpha\beta\delta\gamma}).
\end{equation}
I will consider henceforth the crystal momentum representation, where $\alpha \equiv {\bm k} s$. The electron-electron interaction is taken to be explicitly of the Coulomb form. The many-body Hamiltonian is written as $H^{1e} + V^{ee}$, where
\begin{equation}
\arraycolsep 0.3 ex
\begin{array}{rl}
\displaystyle H^{1e} = & \displaystyle \sum_{{\bm k}s} H_{{\bm k}{\bm k}'ss'} c^\dag_{{\bm k}s} c_{{\bm k}'s'} \\ [3ex]
\displaystyle V^{ee} = & \displaystyle \frac{1}{2} \, \sum_{\bm q} v({\bm q}) \, \sum_{{\bm k}{\bm k}'ss'} c^\dag_{{\bm k} + {\bm q}, s}c^\dag_{{\bm k}' - {\bm q}, s'}c_{{\bm k}'s'}c_{{\bm k}s}.
\end{array}
\end{equation}
The one-particle matrix element $H_{{\bm k}{\bm k}'ss'}$ includes band structure terms and disorder. The matrix element $v({\bm q}) = v(q)$ is given by ($\epsilon_r$ is the relative permittivity)
\begin{equation}
v(q) = \frac{e^2}{2\epsilon_0\epsilon_r q}.
\end{equation}
The real $v({\bm q})$ arises from Coulomb interaction matrix elements between plane waves.
\begin{equation}
\arraycolsep 0.3 ex
\begin{array}{rl}
\displaystyle V_{{\bm k}_1 s_1, {\bm k}_2 s_2, {\bm k}_3 s_3, {\bm k}_4 s_4} = \delta_{s_1s_4} \, \delta_{s_2s_3}\, \delta_{{\bm k_1} + {\bm k}_2, {\bm k}_3 + {\bm k}_4} \, v({\bm k}_2 - {\bm k}_3).
\end{array}
\end{equation}
The term with ${\bm k}_2 = {\bm k}_3$ is canceled by the positive background of the lattice, so $v(0) = 0$.

In TI in the random phase approximation (RPA), abbreviating ${\bm q} = {\bm k}' - {\bm k}$, one replaces $v({\bm q}) \equiv v(q) \rightarrow \displaystyle \frac{v(q)}{\varepsilon(q)}$, where $\varepsilon (q)$ the dielectric function. The polarization function is obtained by summing the lowest bubble diagram. At $T=0$ the long-wavelength limit of the dielectric function is \cite{Hwang_Gfn_Screening_PRB07}
\begin{equation}
\epsilon(q) = 1 + \frac{e^2}{4\pi \epsilon_0 \epsilon_r A} \, \bigg(\frac{k_F}{q}\bigg).
\end{equation}
The polarization function was also calculated in Ref.\ \onlinecite{Raghu_HlcLqd_PRL10}. The screened electron-electron Coulomb potential has the form
\begin{equation}
\displaystyle \frac{v(q)}{\varepsilon(q)} = \frac{e^2}{2 \epsilon_0 \epsilon_r} \frac{1}{\sqrt{k^2 + k^{'2} - 2kk' \cos\gamma_{{\bm k}{\bm k}'}} + \frac{r_sk_F}{2}}.
\end{equation}
The Wigner-Seitz radius $r_s$ (alternatively, the effective fine structure constant), which parametrizes the relative strength of the kinetic energy and electron-electron interactions, is a constant for the Rashba-Dirac Hamiltonian, given by $r_s = e^2/(2\pi \epsilon_0\epsilon_r A)$. In addition to the electron-electron Coulomb potential, the matrix element $\bar{U}_{{\bm k}{\bm k}'}$ of a screened Coulomb potential between plane waves, which will be relevant in transport below, is given by
\begin{equation}\label{eq:W}
\arraycolsep 0.3ex
\begin{array}{rl}
\displaystyle \bar{U}_{{\bm k}{\bm k}'} = & \displaystyle \frac{Ze^2}{2\epsilon_0\epsilon_r}\, \frac{1}{ |{\bm k} - {\bm k}'| + k_{TF}},
\end{array}
\end{equation}
where $Z$ is the ionic charge (which I will assume for simplicity to be $Z = 1$) and $k_{TF} = k_Fr_s/2$ is the Thomas-Fermi wave vector, with $k_F$ the Fermi wave vector.

\section{Density matrix}
\label{sec:DM}

The many-particle density matrix $F$ obeys \cite{Vasko}
\begin{equation}
\td{F}{t} + \frac{i}{\hbar} \, [H, F] = 0.
\end{equation}
The one-particle reduced density matrix $\rho$ is the trace 
\begin{equation}
\rho_{\zeta\eta} = {\rm Tr} \, (c^\dag_\eta c_\zeta F) \equiv \bkt{c^\dag_\eta c_\zeta} \equiv \bkt{F}_{1e}.
\end{equation}
The reduced density matrix satisfies 
\begin{equation}\label{rhoeq}
\arraycolsep 0.3 ex
\begin{array}{rl}
\displaystyle \td{\rho_{\zeta\eta}}{t} + \frac{i}{\hbar} \, [H^{1e}, \rho]_{\zeta\eta} - \frac{i}{\hbar} \, \bkt{[V^{ee}, c^\dag_\eta c_\zeta]} = & \displaystyle 0.
\end{array}
\end{equation}
In terms of the antisymmetric Coulomb matrix element $\tilde{V}_{\alpha\beta\gamma\delta}$ defined above, the last term on the LHS
\begin{equation}
\arraycolsep 0.3 ex
\begin{array}{rl}
\displaystyle [V_{ee}, c^\dag_\eta c_\zeta] = & \displaystyle \sum_{\alpha\beta\gamma} \big[\tilde{V}_{\alpha\beta\gamma\eta} \, c^\dag_\alpha c^\dag_\beta c_\gamma c_\zeta + \tilde{V}_{\beta\gamma\alpha\zeta} \, c^\dag_\eta c^\dag_\alpha c_\beta c_\gamma \big].
\end{array}
\end{equation}
The many-electron average is evaluated as follows
\begin{widetext}
\begin{subequations}
\begin{eqnarray}
\displaystyle \bkt{[V^{ee}, c^\dag_\eta c_\zeta]} & = & \displaystyle \sum_{\alpha\beta\gamma} \bkt{\tilde{V}_{\alpha\beta\gamma\eta} \, c^\dag_\alpha c^\dag_\beta c_\gamma c_\zeta + \tilde{V}_{\beta\gamma\alpha\zeta} \, c^\dag_\eta c^\dag_\alpha c_\beta c_\gamma}  \\
\label{Wick} \displaystyle \bkt{c^\dag_\alpha c^\dag_\beta c_\gamma c_\delta} & = & \displaystyle \bkt{c^\dag_\alpha c_\delta}\bkt{c^\dag_\beta c_\gamma} - \bkt{c^\dag_\alpha c_\gamma} \bkt{c^\dag_\beta c_\delta} + G_{\alpha\beta\gamma\delta}.
\end{eqnarray}
\end{subequations}
The focus of this work is on the first two terms on the RHS of Eq.\ (\ref{Wick}), which represent the Hartree-Fock mean-field part of the electron-electron interaction. The remainder, $G_{\alpha\beta\gamma\delta}$, gives the electron-electron scattering term in the kinetic equation, \cite{Vasko} is second-order in the interaction and vanishes at $T=0$. I will treat the case of zero temperature and reserve electron-electron scattering for a forthcoming publication. To evaluate the Hartree-Fock factorization, note that $\tilde{V}_{\alpha\beta\gamma\eta} \rho_{\gamma\beta}$ cancels, and the remainder becomes
\begin{equation}\label{Average}
\arraycolsep 0.3 ex
\begin{array}{rl}
\displaystyle \bkt{[V_{ee}, c^\dag_\eta c_\zeta]} = & \displaystyle \sum_{\alpha\beta\gamma} \big[\tilde{V}_{\alpha\beta\gamma\eta} \, (\rho_{\zeta\alpha}\rho_{\gamma\beta} - \rho_{\gamma\alpha}\rho_{\zeta\beta}) + \tilde{V}_{\beta\gamma\alpha\zeta} \, (\rho_{\gamma\eta}\rho_{\beta\alpha} - \rho_{\beta\eta}\rho_{\gamma\alpha}) \big].
\end{array}
\end{equation}
\end{widetext}
I will introduce two mean field terms by letting $\tilde{V}_{\alpha\beta\gamma\eta} \rho_{\gamma\beta} = \mathcal{A}^{MF}_{\alpha\eta}$ and $\tilde{V}_{\alpha\beta\gamma\eta} \rho_{\gamma\alpha} = \mathcal{B}^{MF}_{\beta\eta}$, then
\begin{equation}
\arraycolsep 0.3 ex
\begin{array}{rl}
\displaystyle \bkt{[V_{ee}, c^\dag_\eta c_\zeta]} = & \displaystyle [\rho, \mathcal{A}^{MF} - \mathcal{B}^{MF}]_{\zeta\eta}.
\end{array}
\end{equation}
The effective kinetic equation becomes 
\begin{equation}\label{eekineq}
\arraycolsep 0.3 ex
\begin{array}{rl}
\displaystyle \td{\rho}{t} + \frac{i}{\hbar} \, [H_{1e}, \rho] + \frac{i}{\hbar} \, [\mathcal{A}^{MF} - \mathcal{B}^{MF}, \rho] = & \displaystyle 0.
\end{array}
\end{equation}
The one-particle Hamiltonian is renormalized by 
\begin{equation}
H_{ee}^{eff} = \mathcal{A}^{MF} - \mathcal{B}^{MF}.
\end{equation}
I emphasize that in the final analysis one is interested only in the impurity average of $\rho$ in the crystal-momentum representation. In general one may always write $\rho_{{\bm k}{\bm k}'} = f_{\bm k} \delta_{{\bm k}{\bm k}'} + g_{{\bm k}{\bm k}'}$ \cite{Culcer_TI_Kineq_PRB10}, where the ${\bm k}$-off-diagonal part, $g_{{\bm k}{\bm k}'}$, is eventually integrated out to yield the scattering term in any desired approximation. In the impurity average of Eq.\ (\ref{eekineq1}), out of the commutator $\bkt{[H_{ee}^{eff} (\rho), \rho]}_{\zeta\eta}$ only the terms $[H_{ee}^{eff}(f), f]$ and $[H_{ee}^{eff}(g), g]$ survive, where to first order in the electric field $H_{ee}^{eff}(\rho)\rho \equiv H_{ee}^{eff}(\rho_0)\rho_E + H_{ee}^{eff}(\rho_E)\rho_0$. This implies that $H_{ee}^{eff}(g)g \equiv H_{ee}^{eff}(g_0)g_E + H_{ee}^{eff}(g_E)g_0$. In linear response $g_0 = 0$, and we are left with $H_{ee}^{eff}(f)f$. Specializing to this term, spin indices are omitted and $f_{\bm k}$ is treated henceforth as a $2 \times 2$ matrix in spin space.

To determine $H_{ee}^{eff}(f)$, we evaluate the two mean field terms. Beginning with $\mathcal{A}^{MF}$, with summation implied over repeated indices,
\begin{equation}
\arraycolsep 0.3 ex
\begin{array}{rl}
\displaystyle \mathcal{A}^{MF}_{\alpha\eta} = & \displaystyle \tilde{V}_{\alpha\beta\gamma\eta} f_{\gamma\beta} \equiv \tilde{V}_{{\bm k}_1s_1, {\bm k}_2 s_2, {\bm k}_2 s_3, {\bm k}_4s_4} f_{{\bm k}_2s_3s_2} \\ [3ex]
= & \displaystyle v({\bm k}_2 - {\bm k}_3) \, \delta_{s_1s_4} \, \delta_{s_2s_3}\, \delta_{{\bm k_1} + {\bm k}_2, {\bm k}_3 + {\bm k}_4} \, f_{{\bm k}_2s_3s_2} \\ [3ex]
= & \displaystyle v(0) \, \delta_{s_1s_4} \, \delta_{{\bm k_1}{\bm k}_4} \, {\rm tr} \, f_{{\bm k}_2} \\ [3ex]
\therefore & \displaystyle \rightarrow 0, \,\,\, {\rm since} \,\,\, v(0) = 0.
\end{array}
\end{equation}
Therefore $\mathcal{A}^{MF}$ vanishes in the most general case. Next, $\mathcal{B}^{MF}$ is given by
\begin{equation}
\arraycolsep 0.3 ex
\begin{array}{rl}
\displaystyle \mathcal{B}^{MF}_{\beta\eta} = & \displaystyle \tilde{V}_{\alpha\beta\gamma\eta} \rho_{\gamma\alpha} \equiv \sum_{{\bm k}'} v({\bm k} - {\bm k}') \, f_{{\bm k}' s_\beta s_\eta}.
\end{array}
\end{equation}
Note that $\mathcal{B}^{MF}$ can be interpreted as an effective magnetic field due to the Hartree-Fock mean field electron-electron interaction. This result reproduces the correct exchange energy, \cite{Doniach} and yields exchange enhancement of Zeeman field-induced spin polarizations, as found in Fermi liquid theory. It is similar in spirit to the treatment of Ref.~\onlinecite{Tse_Bil_OptCond_PRB09}. 

Equation (\ref{rhoeq}) is reduced to
\begin{equation}\label{eekineq1}
\arraycolsep 0.3 ex
\begin{array}{rl}
\displaystyle \td{\rho}{t} + \frac{i}{\hbar} \, [H^{1e}, \rho] = & \displaystyle \frac{i}{\hbar} \, [\mathcal{B}^{MF}, \rho].
\end{array}
\end{equation}
The single-particle Hamiltonian is renormalized by $\mathcal{B}^{MF}(f)$, which is itself a function of the single-particle density matrix. 

At this stage one may include explicitly disorder and driving electric fields in the one-particle Hamiltonian and write $H^{1e}_{{\bm k}{\bm k}'} = H_{0{\bm k}} \delta_{{\bm k}{\bm k}'} + H_{E{\bm k}{\bm k}'} + U_{{\bm k}{\bm k}'}$, where $H_{0{\bm k}}$ is the band Hamiltonian, $H_{E{\bm k}{\bm k}'} $ the electrostatic potential due to the driving electric field and $U_{{\bm k}{\bm k}'} $ the disorder potential. The effective single-particle kinetic equation takes the form
\begin{widetext}
\begin{equation}\label{eekineq}
\arraycolsep 0.3 ex
\begin{array}{rl}
\displaystyle \td{f_{\bm k}}{t} + \frac{i}{\hbar} \, [H_{0{\bm k}}, f_{\bm k}] + \hat{J} \, (f_{\bm k}) = & \displaystyle - \frac{i}{\hbar} \, [H_{E_{\bm k}}, f_{\bm k}] + \frac{i}{\hbar} \, [\mathcal{B}^{MF}_{\bm k}, f_{\bm k}].
\end{array}
\end{equation}
\end{widetext}
One writes $f_{\bm k} = f_{0{\bm k}} + f_{E{\bm k}} + f^{ee}_{\bm k}$, where $f_{0{\bm k}}$ is the equilibrium density matrix, $f_{E{\bm k}}$ is induced by the electric field, and $f^{ee}_{\bm k}$ by electron-electron interactions \footnote{The interaction correction to the energy contributes a diagonal term to the kinetic equation which drops out of the Hamiltonian and does not contribute to $f_{0{\bm k}}$.}.

Equation (\ref{eekineq}) is solved iteratively in $\mathcal{B}^{MF}$. Let the bare driving term $D_{\bm k} = - \frac{i}{\hbar} \, [H_{E{\bm k}}, f_{\bm k}]$. The approach is to solve the kinetic equation first with $D_{\bm k}$ as the source term. This will give a spin polarization. The spin polarization will give a nonzero $\mathcal{B}^{MF}_{\bm k}$, which in turn will give an additional source term, referred to as $\mathcal{D}^{ee}_{\bm k}$ in the next section. Then one solves the kinetic equation again with $\mathcal{D}^{ee}_{\bm k}$ as the source term
\begin{equation}
\arraycolsep 0.3 ex
\begin{array}{rl}
\displaystyle \td{f_{E{\bm k}}}{t} + \frac{i}{\hbar} \, [H_{0{\bm k}}, f_{E{\bm k}}] + \hat{J} \, (f_{E{\bm k}}) = & \displaystyle - \frac{i}{\hbar} \, [H_{E{\bm k}}, f_{0{\bm k}}] \\ [3ex]
\displaystyle \td{f^{ee}_{\bm k}}{t} + \frac{i}{\hbar} \, [H_{0{\bm k}}, f^{ee}_{\bm k}] + \hat{J} \, (f^{ee}_{\bm k}) = & \displaystyle \frac{i}{\hbar} \, [\mathcal{B}^{MF}_{\bm k}, f_{0{\bm k}}].
\end{array}
\end{equation}
On the RHS of the second equation only the equilibrium density matrix $f_{0{\bm k}}$ appears because $\mathcal{B}^{MF}_{\bm k}$ is first order in the electric field. The iteration is continued to all orders in the Wigner-Seitz radius $r_s$ (that is, to all orders in the effective fine structure constant.) 

We recall that electron-electron and electron-impurity potentials are screened, with screening treated in the random-phase approximation. The density-matrix formalism used here is thus equivalent to the GW approximation. In the non-equilibrium diagram technique, the correction discussed in this work is obtained by including the real part of the Green's function due to electron-electron interactions. \cite{Tse_Bil_OptCond_PRB09} 

\section{Kinetic equation for interacting TI}
\label{sec:KinEq}

Henceforth I specialize to TI. The band Hamiltonian $\displaystyle H_{0{\bm k}} = \displaystyle \frac{\hbar}{2} \, {\bm \sigma} \cdot {\bm \Omega}_{\bm k}$, where ${\bm \Omega}_{\bm k} = \displaystyle - \frac{2Ak}{\hbar} \, \hat{\bm \theta}$, with $\hat{\bm \theta}$ the tangential unit vector in polar coordinates in reciprocal space. Interaction with the electric field is given by $H_{E, {\bm k}{\bm k}'} = \displaystyle (e{\bm E}\cdot\hat{\bm r})_{{\bm k}{\bm k}'} \openone = ie{\bm E}\cdot\pd{}{{\bm k}} \, \delta({\bm k} - {\bm k}') \,  \openone$, with $\openone$ the identity matrix in spin space. Uncorrelated impurities located at ${\bm R}_I$ are represented by $U_{{\bm k}{\bm k}'} = \displaystyle \bar{U}_{{\bm k}{\bm k}'} \sum_I e^{i({\bm k} - {\bm k}')\cdot{\bm R}_I}$, with $\bar{U}_{{\bm k}{\bm k}'}$ the Fourier transform of the potential of a single impurity. I will write $f_{\bm k} = n_{\bm k}\openone + S_{\bm k}$, with $n_{\bm k}$ the number density and $S_{\bm k} = \frac{1}{2}\, {\bm S}_{{\bm k}} \cdot {\bm \sigma}$ the spin density. One decomposes $S_{\bm k} = S_{{\bm k}\parallel} + S_{{\bm k}\perp}$, where $[H_{0{\bm k}}, S_{{\bm k}\parallel}] = 0$ and $S_{{\bm k}\parallel}$ is the fraction of carriers in eigenstates of $H_{0{\bm k}}$, while $S_{{\bm k}\perp}$ represents interband dynamics, i.e. Zitterbewegung. Further, $S_{{\bm k}\parallel} = (1/2) \, s_{{\bm k}\parallel} \, \sigma_{{\bm k} \parallel}$ and $S_{{\bm k}\perp} = (1/2) \, s_{{\bm k}\perp} \sigma_{{\bm k}\perp}$, with the matrices $\sigma_{{\bm k} \parallel} = -{\bm \sigma}\cdot\hat{\bm \theta}$ and $\sigma_{{\bm k}\perp} = {\bm \sigma}\cdot\hat{\bm k}$. 

\subsection{Single-particle kinetic equation}

The general single-particle kinetic equation is
\begin{subequations}\label{eq:Spp}
\begin{eqnarray}
\td{S_{{\bm k} \|}}{t} + P_\| \hat J (S_{{\bm k}}) & = & \mathcal{D}_{{\bm k}\parallel}, \\ [0.5ex]
\td{S_{{\bm k}\perp}}{t} + \frac{i}{\hbar} \, [H_{{\bm k}}, S_{{\bm k}\perp}] + P_\perp \hat J (S_{{\bm k}}) & = & \mathcal{D}_{{\bm k}\perp},
\end{eqnarray}
\end{subequations}
where the driving term $\mathcal{D}_{{\bm k}} = \displaystyle \frac{e{\bm E}}{\hbar} \cdot \pd{\rho_{0{\bm k}}}{\bm k}$, and $\rho_{0{\bm k}}$ is the equilibrium density matrix. This equation is solved as an expansion in the small parameter $\hbar/(\varepsilon_F\tau)$, where the momentum relaxation time $\tau$ is defined below. The leading-order term in this expansion is $\propto [\hbar/(\varepsilon_F\tau)]^{(-1)}$ and is found from
\begin{equation}\label{eq:Spert}
P_\| \hat J (S_{{\bm k}\parallel}) = \mathcal{D}_{{\bm k}\parallel}.
\end{equation}

The solution to Eq.\ (\ref{eq:S}) requires certain approximations. With respect to the scattering potential one expands in the small parameter $\hbar/(\varepsilon_F\tau)$. In the steady state in the Born approximation the leading term in the solution to the kinetic equation is $\propto [\hbar/(\varepsilon_F\tau)]^{(-1)}$. It is trivial to check that at finite doping the next term in the expansion, i.e. $\propto [\hbar/(\varepsilon_F\tau)]^{(0)}$, vanishes identically, which was demonstrated in Ref\ \onlinecite{Culcer_TI_Kineq_PRB10}. A term $\propto [\hbar/(\varepsilon_F\tau)]^{(0)}$ would appear in the weak localization regime, yet this correction is not relevant in the regime $\varepsilon_F\tau/\hbar \gg 1$ considered in this work.

The Born-approximation scattering term has the form
\begin{equation}
\label{JBorn}
\hat{J}(f_{\bm k}) = \frac{1}{\hbar^2} \bigg\langle\bigg\langle\int_0^{\infty} dt' \, [\hat U, e^{- \frac{i \hat H t'}{\hbar}}[\hat U, \hat f]\, e^{ \frac{i \hat H t'}{\hbar}}]\bigg\rangle\bigg\rangle_{{\bm k}{\bm k}},
\end{equation}
with $\gamma = \theta' - \theta$ the angle between the incident and scattered wave vectors, ${\bm k}$ and ${\bm k}'$ respectively, and $\bkt{\bkt{}}$ denoting the average over impurity configurations. The projections of $\hat{J}(f_{\bm k})$ needed in this work have been determined before \cite{Culcer_TI_Kineq_PRB10}
\begin{equation}
\arraycolsep 0.3ex
\begin{array}{rl}
\displaystyle P_\parallel \hat J(S_{{\bm k}\parallel}) = & \displaystyle \frac{kn_i\, \sigma_{{\bm k}\parallel}}{8\hbar \pi A} \int d\gamma \, |\bar{U}_{{\bm k}{\bm k}'}|^2 \, (s_{{\bm k}\parallel} - s_{{\bm k}'\parallel})(1 + \cos\gamma) \\ [3ex]
\displaystyle P_\perp \hat J(S_{{\bm k}\parallel}) = & \displaystyle \frac{kn_i\, \sigma_{{\bm k} \perp} }{8\hbar \pi A} \int\! d\gamma \, |\bar{U}_{{\bm k}{\bm k}'}|^2\, (s_{{\bm k}\parallel} - s_{{\bm k}'\parallel}) \sin\gamma \\ [3ex]
\displaystyle P_\parallel \hat J(S_{{\bm k}\perp}) = & \displaystyle \frac{kn_i \, \sigma_{{\bm k}\parallel}}{8\hbar \pi A} \int\! d\gamma \, |\bar{U}_{{\bm k}{\bm k}'}|^2\, \big(s_{{\bm k}\perp} + s_{{\bm k}'\perp}\big) \sin\gamma,
\end{array}
\end{equation} 
where $\gamma = \theta' - \theta$ is the angle between the incident and scattered wave vectors. The small $s_{{\bm k}\parallel}$ and $s_{{\bm k}\perp}$ are scalars, $s_{{\bm k}\parallel} = - {\bm S}_{\bm k} \cdot\hat{\bm \theta}$ and $s_{{\bm k}\perp} = {\bm S}_{\bm k} \cdot\hat{\bm k}$. 

The scattering terms contain factors of $(1 + \cos\gamma)$ (reflecting the $\pi$ Berry phase) or $\sin\gamma$, both of which prohibit backscattering and give rise to Klein tunneling. Since the current operator $\propto {\bm \sigma}$, only $S_{\bm k}$ is needed. In the absence of scalar terms in the Hamiltonian, $\hat{J}(f_{\bm k})$ does not couple $n_{\bm k}$ with $S_{\bm k}$, and Eq.\ (\ref{eekineq}) makes evident the fact that the interaction term does not couple $n_{\bm k}$ and $S_{\bm k}$, thus $n_{\bm k}$ may be dispensed with for the remainder of this work. The equation satisfied by $S_{\bm k}$ is
\begin{widetext}
\begin{equation}\label{eq:S}
\arraycolsep 0.3 ex
\begin{array}{rl}
\displaystyle \td{S_{\bm k}}{t} + \frac{i}{\hbar} \, [H_{0{\bm k}}, S_{\bm k}] + \hat{J} \, (S_{\bm k}) = & \displaystyle - \frac{i}{\hbar} \, [H_{E_{\bm k}}, S_{\bm k}]  + \frac{i}{\hbar} \, [\mathcal{B}^{MF, (1)}_{\bm k}, S_{\bm k}].
\end{array}
\end{equation}
\end{widetext}
With respect to the electron-electron interaction one also needs to define a perturbation expansion in order to solve Eq.\ (\ref{eq:S}), which is done in what follows. Within the approximations used in this paper, this expansion can be summed \textit{exactly}. The method of solution is summarized as follows. The kinetic equation first with $\mathcal{B}^{MF}_{\bm k}$ set to zero. This solution is already known \cite{Culcer_TI_Kineq_PRB10} and gives a spin polarization, which in turn generates a nonzero $\mathcal{B}^{MF}_{\bm k}$, which itself yields a new driving term, and so forth. The full solution is found as a perturbation expansion in the electron-electron interaction, which can be summed exactly. 

To obtain the solution in the interacting case, it is therefore first necessary to solve the non-interacting problem. In the absence of interactions \cite{Culcer_TI_Kineq_PRB10} the steady-state solution to the density matrix in the Born approximation is \cite{Culcer_TI_Kineq_PRB10} 
\begin{equation}\label{Spar}
\arraycolsep 0.3ex
\begin{array}{rl}
\displaystyle S_{E{\bm k} \|} = & \displaystyle \frac{\tau\, e \,{\bm E}\cdot\hat{\bm k}}{4\hbar} \, \pd{f_{0+}}{k} \, \sigma_{{\bm k} \|} \\ [2ex]
\displaystyle \frac{1}{\tau} = & \displaystyle \frac{kn_i}{4\hbar A}\int \frac{d\gamma}{2\pi} \, |\bar{U}_{{\bm k}{\bm k}'}|^2\sin^2\gamma.
\end{array} 
\end{equation}
Above $n_i$ is the impurity density, while the factor of $\sin^2\gamma$ represents the product $(1 + \cos\gamma)(1 - \cos\gamma)$. The first term in this product is characteristic of TI and ensures backscattering is suppressed, while the second term is characteristic of transport, eliminating the effect of small-angle scattering. In non-interacting TI the Zitterbewegung contribution to the conductivity/spin-density (i.e. due to $S_{E{\bm k} \perp}^{ee, (0)}$) vanishes identically at finite doping. But in the interacting case it is necessary to consider both the electron and the hole bands to capture the spin dynamics. 

The solution obtained, $S_{E{\bm k}} \equiv S_{E{\bm k} \|}$, is fed into $\mathcal{B}^{MF}_{\bm k}$, which in turn generates a new driving term in the kinetic equation. Each term in this expansion by the index $\alpha$, thus $\mathcal{B}^{MF, (\alpha)}_{\bm k}$. The solution found in Eq.\ \ref{Spar} corresponds to $\alpha = 0$, that is, in the non-interacting case $S_{E{\bm k} \|} \equiv S_{E{\bm k} \|}^{ee, (0)}$. The driving term due to electron-electron interactions is generically denoted $\mathcal{D}_{{\bm k}}^{ee, (\alpha)}$. The decomposition $\mathcal{D}_{{\bm k}\perp}^{ee, (\alpha)} = (1/2) \, d_{{\bm k}\perp}^{ee, (\alpha)} \sigma_{{\bm k}\perp}$ is also used. The solution to the spin part of the density matrix to order $\alpha$ is denoted by $S_{{\bm k}}^{ee, (\alpha)}$. The driving term $\mathcal{D}_{{\bm k}}^{ee, (\alpha)}$ is \textit{always} orthogonal to $H_{0{\bm k}}$, therefore $\mathcal{D}_{{\bm k}}^{ee, (\alpha)} \equiv \mathcal{D}_{{\bm k}\perp}^{ee, (\alpha)}$. The kinetic equation for the solution $S^{ee}_{\bm k}$ in the presence of electron-electron interactions can be written for each order as 
\begin{subequations}\label{eq:Slead}
\begin{eqnarray}\label{eq:Slead1}
\td{S_{{\bm k}\perp}^{ee, (\alpha)}}{t} + \frac{i}{\hbar} \, [H_{{\bm k}}, S_{{\bm k}\perp}^{ee, (\alpha)}] & = & \mathcal{D}_{{\bm k}\perp}^{ee, (\alpha)} \\ [0.5ex]\label{eq:Slead2}
P_\| \hat J (S_{{\bm k}}^{ee, (\alpha)}) & = & - P_\| \hat J (S_{{\bm k}\perp}^{ee, (\alpha)}), 
\end{eqnarray}
\end{subequations}
where in Eq.\ (\ref{eq:Slead2}) it is understood that the RHS, found from Eq.\ (\ref{eq:Slead1}), acts as the source for the LHS. The scattering term does not appear in Eq.\ (\ref{eq:Slead1}) since, as was argued above, $\mathcal{D}_{{\bm k}\parallel}^{ee, (\alpha)} = 0$ always. 

I will dwell first on the solution of Eq.\ (\ref{eekineq}) due to $\mathcal{B}^{MF, (1)}_{\bm k}$, i.e. first order in the interaction, which requires $S_{E{\bm k} \|}^{ee, (0)}$. From Eq.\ (\ref{Spar}),
\begin{equation}\label{Hee1}
\displaystyle \mathcal{B}^{MF, (1)}_{\bm k} = \frac{e^2 k_F}{4\pi \epsilon_0 \epsilon_r} \int_0^1 \!\!\! dl' \, l' \!\!\! \int_0^{2\pi}\!\!\! \frac{d\gamma}{2\pi} \frac{S_{E{\bm k}'}^{ee, (0)}}{\sqrt{l^2 + l^{'2} - 2ll' \, \cos\gamma} + \frac{r_s}{2}},
\end{equation}
where $l = (k/k_F)$. The term in $\mathcal{B}^{MF, (1)}_{\bm k}$ in which $S_{{\bm k}'} \rightarrow S_{0{\bm k}'}$ gives a vanishing contribution. For ${\bm E} \parallel \hat{\bm x}$
\begin{equation}
\arraycolsep 0.3 ex
\begin{array}{rl}
\displaystyle \mathcal{B}^{MF, (1)}_{\bm k} = & \displaystyle \frac{e^3E_x\tau}{16 \pi \hbar \epsilon_0 \epsilon_r} \, [I_c^{(1)} \cos\theta \, \sigma_{{\bm k} \|} - I_{ee}^{(1)} \sin\theta \, \sigma_{{\bm k}\perp}] \\ [1ex]
\displaystyle I_{ee}^{(1)}(l, r_s) = & \displaystyle \frac{1}{2}\int \frac{d\gamma}{2\pi} \, \frac{(\cos2\gamma - 1)}{(\sqrt{1 + l^2 - 2l \, \cos\gamma} + \frac{r_s}{2}}. 
\end{array}
\end{equation}
In $I_c^{(1)}(l, r_s)$ the sign of the cosine term is flipped. Although $\mathcal{B}^{MF, (1)}_{\bm k}$ itself has a part $\propto \sigma_{{\bm k} \|}$, this part drops out of the driving term in Eq.\ (\ref{eq:Seeperp}), because one is working to first order in the electric field and the commutator $[\mathcal{B}^{MF, (1)}_{\bm k}, S_{{\bm k}}] \rightarrow [\mathcal{B}^{MF, (1)}_{\bm k}, S_{0{\bm k}}]$, and $S_{0{\bm k}} \propto \sigma_{{\bm k} \|}$. The effective electron-electron interaction Hamiltonian $\mathcal{B}^{MF, (1)}_{\bm k}$ therefore contributes a driving term orthogonal to $H_{0{\bm k}}$, yielding a correction $S^{ee, (1)}_{E{\bm k}\perp}$ to the density matrix. The scattering term does not appear in the equation for $S_{{\bm k}\perp}^{ee, (1)}$. Equation (\ref{eq:S}) takes the simple form
\begin{equation}\label{eq:Seeperp}
\td{S^{ee, (1)}_{E{\bm k}\perp}}{t} + \frac{i}{\hbar} \, [H_0, S^{ee, (1)}_{E{\bm k}\perp}] = \frac{i}{\hbar} \, [\mathcal{B}^{MF, (1)}_{\bm k}, S_{0{\bm k}}].
\end{equation}
This equation is solved using the time evolution operator
\begin{equation}\label{eq:Sperp1}
S^{ee, (1)}_{E{\bm k}\perp} = \displaystyle \frac{eE_x r_s \tau I_{ee}^{(1)} (l, r_s)}{16 \hbar k} \, f_{0+}\sin\theta \, {\bm \sigma}\cdot \hat{\bm k}.
\end{equation}
Another contribution stems from the projection
\begin{equation}
\arraycolsep 0.3ex
\begin{array}{rl}
\displaystyle P_\parallel \hat J[S^{ee, (1)}_{{\bm k}\parallel}] = & \displaystyle - P_\parallel \hat J[S^{ee, (1)}_{{\bm k}\perp}].
\end{array}
\end{equation}
It is understood that the RHS, found from Eq.\ (\ref{eq:Sperp1}), acts as the source for the LHS. Straightforwardly
\begin{equation}
\arraycolsep 0.3ex
\begin{array}{rl}\label{See1parallel}
\displaystyle S^{ee, (1)}_{{\bm k}\parallel} = & \displaystyle \frac{eE_x r_s \tau I_{ee}^{(1)} (l, r_s)}{16 \hbar k} \, f_{0+} \cos\theta \, {\bm \sigma} \cdot \hat{\bm \theta}.
\end{array}
\end{equation}
$S^{ee, (1)}_{E{\bm k}\perp}$ and $S^{ee, (1)}_{{\bm k}\parallel}$ contribute equally to the charge current determined below. In effect, scattering from $S^{ee, (1)}_{E{\bm k}\perp}$ into $S^{ee, (1)}_{{\bm k}\parallel}$ doubles the contribution to the electrical conductivity due to $S^{ee, (1)}_{E{\bm k}\perp}$.
 
The longitudinal current density operator $j_x = \displaystyle \frac{eA}{\hbar} \, \sigma_y$: the current density is equivalent to a spin polarization. The conductivity $\sigma_{xx}^0$ of the non-interacting system is $\sigma_{xx}^0 = \displaystyle \bigg(\frac{e^2}{h}\bigg) \, \bigg( \frac{Ak_F\tau}{4\hbar} \bigg)$ \cite{Culcer_TI_Kineq_PRB10}. The first-order conductivity correction in the electron-electron interaction is
\begin{equation}\label{eq:sigmaee1}
\arraycolsep 0.3 ex
\begin{array}{rl}
\displaystyle \sigma^{ee, (1)}_{xx} = & \displaystyle \bigg(\frac{e^2}{h}\bigg) \, \bigg(\frac{Ak_F\tau}{4\hbar}\bigg) \, \frac{r_sI_{ee}^{(1)}}{2} \equiv \sigma_{xx}^0 \, \bigg(\frac{r_sI_{ee}^{(1)}}{2}\bigg)
\end{array}
\end{equation}
where $I_{ee}^{(1)}(r_s) = \displaystyle \int_0^1 dl \, I_{ee}^{(1)} (l, r_s)$. The electrical current and non-equilibrium spin polarization are renormalized by electron-electron interactions.

Equation\ (\ref{eq:sigmaee1}) has been obtained to first order in the (screened) interaction. The source term due to $d_{E {\bm k}}^{ee, (1)}$ contains only $e^{\pm i\theta}$, identical in structure to the non-interacting problem \cite{Culcer_TI_Kineq_PRB10}. One solves for all higher terms in $H_{\bm k}^{ee}$ by iterating steps (\ref{Hee1})-(\ref{See1parallel}), obtaining the exact result for the conductivity (and spin polarization)
\begin{equation}\label{eq:sigmaeeall}
\arraycolsep 0.3 ex
\begin{array}{rl}
\displaystyle \frac{\sigma_{xx}}{\sigma_{xx}^0} = & \displaystyle 1 + \frac{r_s}{2}\, \bigg[I_{ee}^{(1)} + \frac{r_s}{4}\, I_{ee}^{(2)} + \bigg(\frac{r_s}{4}\bigg)^2 \, I_{ee}^{(3)} + ... \bigg].
\end{array}
\end{equation}
The general formula for the dimensionless integral $I_{ee}^{(n)}$ for $n >1$ is
\begin{widetext}
\begin{equation}
\arraycolsep 0.3ex
\begin{array}{rl}
\displaystyle I_{ee}^{(n)} = & \displaystyle \Pi_{i=1}^{i=n} \int_0^1 dl_i \int_0^{2\pi} \frac{d\gamma_i}{2\pi} \, \bigg(\frac{1}{\frac{r_s}{2} + \sqrt{l_1^2 + l_2^2 - 2l_1l_2 \cos\gamma_1}}\bigg) \bigg(\frac{1}{\frac{r_s}{2} + \sqrt{l_2^2 + l_3^2 - 2l_2l_3 \cos\gamma_2}}\bigg)...\bigg( \frac{(-1)^n\sin^2\gamma_n}{\frac{r_s}{2} + \sqrt{1 + l_n^2 - 2l_n \cos\gamma_n}}\bigg).
\end{array}
\end{equation} 
\end{widetext}

In 2D $v(q) \propto 1/q$, while in TI $r_s$ is density-independent, and the screened Coulomb potential $v(q)/\epsilon(q)\propto 1/q$ also. Thus $H_{\bm k}^{ee}$ does not introduce density dependence: at larger densities the Coulomb interaction is weaker. Solving for $S^{ee}_{E{\bm k}\perp}$ introduces a factor of $1/\Omega_{\bm k}$, which is canceled by $k$ in the 2D volume element. Thus, 2D physics and TI linear dispersion combine to ensure the renormalization is density independent. 

The renormalization reflects the interplay of spin-momentum locking and many-body correlations. A spin at ${\bm k}$ feels the effect of two competing interactions. The Coulomb interaction between Bloch electrons with ${\bm k}$ and ${\bm k}'$ tends to align a spin at ${\bm k}$ with the spin at ${\bm k}'$, equivalent to a $\hat{\bm z}$-rotation -- hence the driving term in Eq.\ (\ref{eq:Seeperp}) is $\propto \sigma_z$. The total mean-field interaction tends to align the spin at ${\bm k}$ with the sum of all spins at all ${\bm k}'$, and $H^{ee}_{\bm k}$ encapsulates the amount by which the spin at ${\bm k}$ is tilted as a result of the mean-field interaction with all other spins on the Fermi surface. The effective field ${\bm \Omega}_{\bm k}$ tends to align the spin with itself. As a result of this latter fact, out of equilibrium, an electrically-induced spin polarization is already found in the non-interacting system \cite{Culcer_TI_Kineq_PRB10}. Given that the spins at ${\bm k}$ and ${\bm k}'$ are in the plane, interactions tilt the spin at ${\bm k}$ in the direction of the spin at ${\bm k}'$. Thus far the argument helps one understand why, if there is no spin polarization to start with, electron-electron interactions do not give rise to a spin polarization. The mean-field result is zero, so there is no overall tilt on any one spin due to the spins of the remaining electrons. Interactions tend to align electron spins in the direction of the existing polarization. The effective $\hat{\bm z}$-rotation explains the counterintuitive observation that the renormalization is related to interband dynamics, originating as it does in $S_{{\bm k}\perp}$. Many-body interactions give an effective ${\bm k}$-dependent magnetic field $\parallel\hat{\bm z}$, such that for ${\bm E} \parallel \hat{\bm x}$ the spins $s_y$ and $-s_y$ are rotated in opposite directions. Due to spin-momentum locking, a tilt in the spin becomes a tilt in the wave vector: spin dynamics create a feedback effect on charge transport, renormalizing the conductivity. This feedback effect is even clearer in the fact that the projection $-P_\parallel \hat J(S_{{\bm k}\perp})$ doubles the renormalization. This doubling is valid for \textit{any} elastic scattering. 

\begin{figure}[tbp]
\bigskip
\includegraphics[width=\columnwidth]{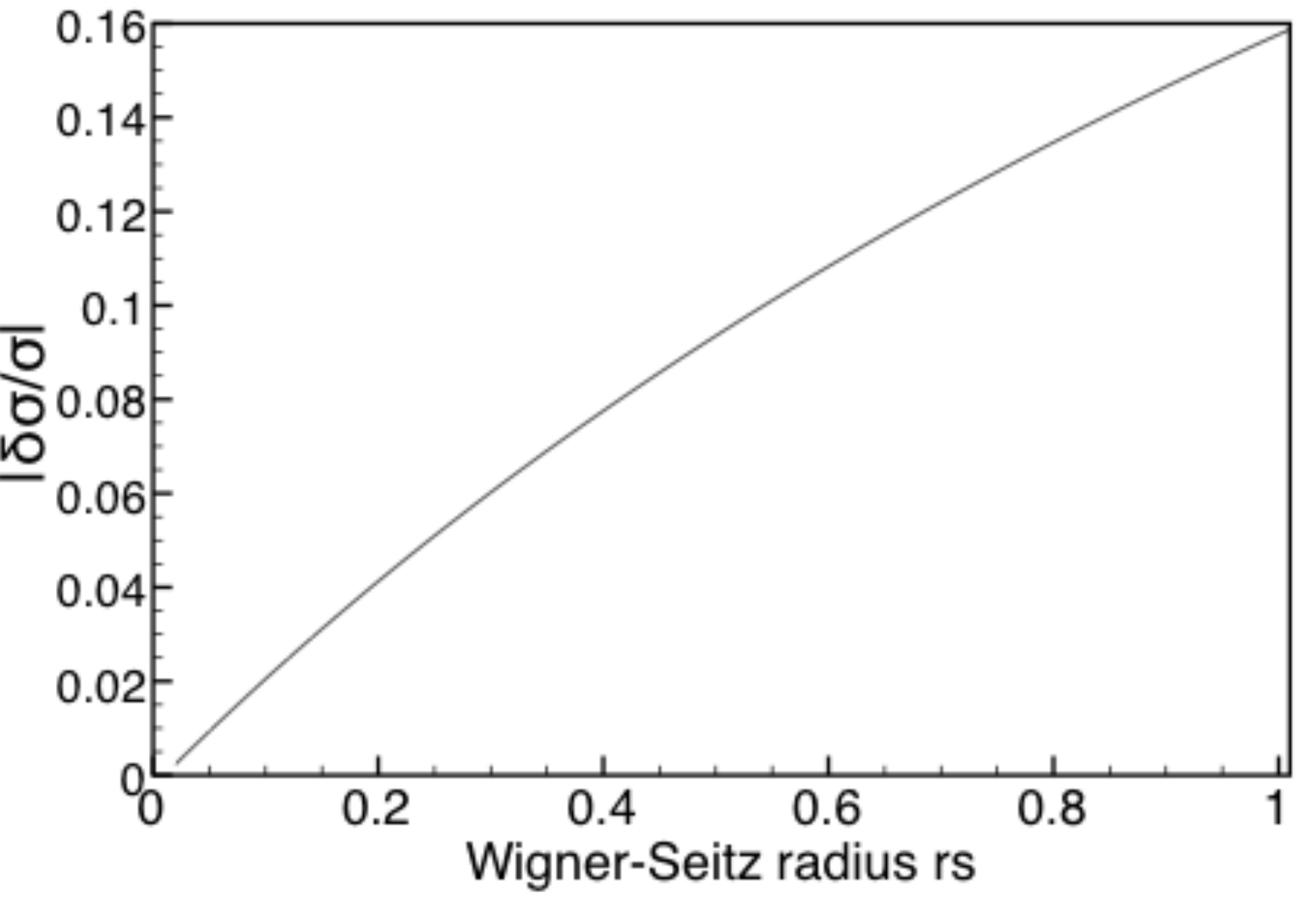}
\caption{\label{rsInt} Fractional change in the conductivity $|\delta\sigma^{ee}_{xx}/\sigma_{xx}^0|$ for $0 \le r_s \le 1$. The current generation of topological insulators has $r_s \ll 1$, so the theory presented in this work is applicable to these materials.}
\end{figure}


I will discuss next the magnitude of this renormalization in currently known topological insulators. Several materials were predicted to be topological insulators in three dimensions. The first was the alloy Bi$_{1-x}$Sb$_x$,\cite{Teo_BiSb_SfcStt_PRB08, Zhang_BiSb_SfcStt_PRB09} followed by the tetradymite semiconductors Bi$_2$Se$_3$, Bi$_2$Te$_3$ and Sb$_2$Te$_3$. \cite{Zhang_TI_BandStr_NP09} These materials have a rhombohedral structure composed of quintuple layers oriented perpendicular to the trigonal $c$-axis. The covalent bonding within each quintuple layer is much stronger than weak van der Waals forces bonding neighboring layers. The semiconducting gap is approximately 0.3 eV, and the TI states are present along the (111) direction. In particular Bi$_2$Se$_3$ and Bi$_2$Te$_3$ have long been known from thermoelectric transport as displaying sizable Peltier and Seebeck effects, and their high quality has ensured their place at the forefront of experimental attention. \cite{Hasan_TI_RMP10} Initial predictions of the existence of chiral surface states were confirmed by first principles studies of Bi$_2$Se$_3$, Bi$_2$Te$_3$, and Sb$_2$Te$_3$. \cite{Zhang_TI_1stPrinc_NJP10} In the current generation of topological insulators, $r_s$ is small. Currently $\epsilon_r$ ranges between 30 and 100 (200 for Bi$_2$Te$_3$), making $r_s$ between 0.14 and 0.46. The theoretical treatment adopted in this work is therefore applicable, and interactions provide a correction to the steady-state response. A plot of $|\delta\sigma^{ee}_{xx}/\sigma_{xx}^0|$ for $0 \le r_s \le 1$ is shown in Fig.~\ref{rsInt}, from which it emerges that interactions may account for up to $\approx 15\%$ of the observed conductivity of surface states in the regime studied. I note that Heusler alloys were recently predicted to have topological surface states, \cite{Xiao_Heusler_PRL10} as well as chalcopyrites, \cite{Feng_TI_Chalcop_PRL11} yet more work is needed to establish the size of $r_s$ in these materials. 

At this stage in topological insulator research, the results found in this work are interesting for conceptual reasons, since they demonstrate that TI phenomenology is unchanged by interactions. The electrical conductivity/spin polarization has the same form as in the non-interacting case, with a renormalization that can be incorporated into a redefinition of the spin-orbit constant, or alternatively of the Fermi velocity, and thus the density of states. For large $r_s$ a non-perturbative treatment that goes beyond the random phase approximation is necessary, yet such a theory must await materials progress. In this context, I would like to note that the growth of new materials is a nontrivial issue, and obtaining high-quality samples where only the surface electrons can be accessed in transport has proved to be a difficult task. It is especially important to recall that future work may initially be hampered by factors such as roughness and dirt inherent in solid-state interfaces. In addition, it remains true that all current TI materials are effectively bulk \textit{metals} because of their large unintentional doping -- at present, bulk carriers are only removed temporarily by gating. Discussing TI surface transport in such bulk-doped TI materials retains some ambiguity, since it necessarily involves complex data fitting and a series of assumptions required by the necessity of distinguishing bulk versus surface transport contributions. Real progress is expected when surface TI transport can be carried out unambiguously, without any complications arising from the (more dominant) bulk transport channel. The immediate tasks facing experimentalists are getting the chemical potential in the gap without the aid of a gate, further experimental studies confirming ambipolar transport, and the measurement of a spin-polarized current. 

\section{Conclusions}

I have demonstrated that, from the point of view of the non-equilibrium spin polarizations and charge current, TI behavior remains intact in the presence of interactions with only quantitative modifications. The conductivity and spin polarization are renormalized by electron-electron interactions entering through a combination of interband dynamics and scattering. 

I am greatly indebted to S. Das Sarma, A.~H.~MacDonald, Yafis Barlas, W.~K.~Tse, Stephen Powell, Junren Shi, Jeil Jung, Dagim Tilahun, and Arun Paramekanti. This work was supported in part by the Chinese Academy of Sciences and in part by the National Science Foundation under Grant No. NSF PHY05-51164. 


\begin{thebibliography}{47}
\expandafter\ifx\csname natexlab\endcsname\relax\def\natexlab#1{#1}\fi
\expandafter\ifx\csname bibnamefont\endcsname\relax
  \def\bibnamefont#1{#1}\fi
\expandafter\ifx\csname bibfnamefont\endcsname\relax
  \def\bibfnamefont#1{#1}\fi
\expandafter\ifx\csname citenamefont\endcsname\relax
  \def\citenamefont#1{#1}\fi
\expandafter\ifx\csname url\endcsname\relax
  \def\url#1{\texttt{#1}}\fi
\expandafter\ifx\csname urlprefix\endcsname\relax\def\urlprefix{URL }\fi
\providecommand{\bibinfo}[2]{#2}
\providecommand{\eprint}[2][]{\url{#2}}

\bibitem[{\citenamefont{Kane and Mele}(2005)}]{KaneMele_QSHE_PRL05}
\bibinfo{author}{\bibfnamefont{C.~L.} \bibnamefont{Kane}} \bibnamefont{and}
  \bibinfo{author}{\bibfnamefont{E.~J.} \bibnamefont{Mele}},
  \bibinfo{journal}{Phys.\ Rev.\ Lett.} \textbf{\bibinfo{volume}{95}},
  \bibinfo{pages}{226801} (\bibinfo{year}{2005}).

\bibitem[{\citenamefont{Hasan and Kane}(2010)}]{Hasan_TI_RMP10}
\bibinfo{author}{\bibfnamefont{M.~Z.} \bibnamefont{Hasan}} \bibnamefont{and}
  \bibinfo{author}{\bibfnamefont{C.~L.} \bibnamefont{Kane}},
  \bibinfo{journal}{Rev.\ Mod.\ Phys.} \textbf{\bibinfo{volume}{82}},
  \bibinfo{pages}{3045} (\bibinfo{year}{2010}).

\bibitem[{\citenamefont{Qi and Zhang}(2011)}]{Qi_TI_RMP_10}
\bibinfo{author}{\bibfnamefont{X.-L.} \bibnamefont{Qi}} \bibnamefont{and}
  \bibinfo{author}{\bibfnamefont{S.-C.} \bibnamefont{Zhang}},
  \bibinfo{journal}{Rev.\ Mod.\ Phys.} \textbf{\bibinfo{volume}{83}},
  \bibinfo{pages}{1057} (\bibinfo{year}{2011}).

\bibitem[{\citenamefont{Moore and
  Balents}(2007)}]{Moore_TRI_TI_Invariants_PRB07}
\bibinfo{author}{\bibfnamefont{J.~E.} \bibnamefont{Moore}} \bibnamefont{and}
  \bibinfo{author}{\bibfnamefont{L.}~\bibnamefont{Balents}},
  \bibinfo{journal}{Phys.\ Rev.\ B} \textbf{\bibinfo{volume}{75}},
  \bibinfo{pages}{121306} (\bibinfo{year}{2007}).

\bibitem[{\citenamefont{Bychkov and Rashba}(1984)}]{BychkovRashba_JETP84}
\bibinfo{author}{\bibfnamefont{Y.~A.} \bibnamefont{Bychkov}} \bibnamefont{and}
  \bibinfo{author}{\bibfnamefont{E.~I.} \bibnamefont{Rashba}},
  \bibinfo{journal}{JETP Lett.} \textbf{\bibinfo{volume}{39}},
  \bibinfo{pages}{66} (\bibinfo{year}{1984}).

\bibitem[{\citenamefont{Zang and
  Nagaosa}(2010)}]{ZangNagaosa_TI_Monopole_PRB10}
\bibinfo{author}{\bibfnamefont{J.}~\bibnamefont{Zang}} \bibnamefont{and}
  \bibinfo{author}{\bibfnamefont{N.}~\bibnamefont{Nagaosa}},
  \bibinfo{journal}{Phys.\ Rev.\ B} \textbf{\bibinfo{volume}{81}},
  \bibinfo{pages}{245125} (\bibinfo{year}{2010}).

\bibitem[{\citenamefont{Tse and MacDonald}(2010)}]{Tse_TI_GiantMOKE_PRL10}
\bibinfo{author}{\bibfnamefont{W.-K.} \bibnamefont{Tse}} \bibnamefont{and}
  \bibinfo{author}{\bibfnamefont{A.~H.} \bibnamefont{MacDonald}},
  \bibinfo{journal}{Phys.\ Rev.\ Lett.} \textbf{\bibinfo{volume}{105}},
  \bibinfo{pages}{057401} (\bibinfo{year}{2010}).

\bibitem[{\citenamefont{Fu and Kane}(2008)}]{Fu_Proximity_Majorana_PRL08}
\bibinfo{author}{\bibfnamefont{L.}~\bibnamefont{Fu}} \bibnamefont{and}
  \bibinfo{author}{\bibfnamefont{C.}~\bibnamefont{Kane}},
  \bibinfo{journal}{Phys.\ Rev.\ Lett.} \textbf{\bibinfo{volume}{100}},
  \bibinfo{pages}{096407} (\bibinfo{year}{2008}).

\bibitem[{\citenamefont{Nayak et~al.}(2008)\citenamefont{Nayak, Simon, Stern,
  Freedman, and {Das Sarma}}}]{SDS_TQC_RMP08}
\bibinfo{author}{\bibfnamefont{C.}~\bibnamefont{Nayak}},
  \bibinfo{author}{\bibfnamefont{S.~H.} \bibnamefont{Simon}},
  \bibinfo{author}{\bibfnamefont{A.}~\bibnamefont{Stern}},
  \bibinfo{author}{\bibfnamefont{M.}~\bibnamefont{Freedman}}, \bibnamefont{and}
  \bibinfo{author}{\bibfnamefont{S.}~\bibnamefont{{Das Sarma}}},
  \bibinfo{journal}{Rev.\ Mod.\ Phys.} \textbf{\bibinfo{volume}{80}},
  \bibinfo{pages}{1083} (\bibinfo{year}{2008}).

\bibitem[{\citenamefont{Doniach and Sondheimer}(1998)}]{Doniach}
\bibinfo{author}{\bibfnamefont{S.}~\bibnamefont{Doniach}} \bibnamefont{and}
  \bibinfo{author}{\bibfnamefont{E.~H.} \bibnamefont{Sondheimer}},
  \emph{\bibinfo{title}{Green's Functions for Solid State Physicists}}
  (\bibinfo{publisher}{Imperial College Press}, \bibinfo{address}{London, UK},
  \bibinfo{year}{1998}).

\bibitem[{\citenamefont{Levin and Stern}(2009)}]{Levin_FracTI_PRL09}
\bibinfo{author}{\bibfnamefont{M.}~\bibnamefont{Levin}} \bibnamefont{and}
  \bibinfo{author}{\bibfnamefont{A.}~\bibnamefont{Stern}},
  \bibinfo{journal}{Phys.\ Rev.\ Lett.} \textbf{\bibinfo{volume}{103}},
  \bibinfo{pages}{196803} (\bibinfo{year}{2009}).

\bibitem[{\citenamefont{Wu et~al.}(to be published)\citenamefont{Wu, Rachel,
  Liu, and Hur}}]{Wu_QSH_Interact_11}
\bibinfo{author}{\bibfnamefont{W.}~\bibnamefont{Wu}},
  \bibinfo{author}{\bibfnamefont{S.}~\bibnamefont{Rachel}},
  \bibinfo{author}{\bibfnamefont{W.-M.} \bibnamefont{Liu}}, \bibnamefont{and}
  \bibinfo{author}{\bibfnamefont{K.~L.} \bibnamefont{Hur}},
  \bibinfo{journal}{arXiv:1106.0943}  (\bibinfo{year}{to be published}).

\bibitem[{\citenamefont{Varney et~al.}(2010)\citenamefont{Varney, Sun, Rigol,
  and Galitski}}]{Varney_TI_ee_QPT_PRB10}
\bibinfo{author}{\bibfnamefont{C.~N.} \bibnamefont{Varney}},
  \bibinfo{author}{\bibfnamefont{K.}~\bibnamefont{Sun}},
  \bibinfo{author}{\bibfnamefont{M.}~\bibnamefont{Rigol}}, \bibnamefont{and}
  \bibinfo{author}{\bibfnamefont{V.}~\bibnamefont{Galitski}},
  \bibinfo{journal}{Phys.\ Rev.\ B} \textbf{\bibinfo{volume}{82}},
  \bibinfo{pages}{115125} (\bibinfo{year}{2010}).

\bibitem[{\citenamefont{Seradjeh
  et~al.}(2009{\natexlab{a}})\citenamefont{Seradjeh, Moore, and
  Franz}}]{Seradjeh_TI_XtnCond_PRL09}
\bibinfo{author}{\bibfnamefont{B.}~\bibnamefont{Seradjeh}},
  \bibinfo{author}{\bibfnamefont{J.~E.} \bibnamefont{Moore}}, \bibnamefont{and}
  \bibinfo{author}{\bibfnamefont{M.}~\bibnamefont{Franz}},
  \bibinfo{journal}{Phys.\ Rev.\ Lett.} \textbf{\bibinfo{volume}{103}},
  \bibinfo{pages}{066402} (\bibinfo{year}{2009}{\natexlab{a}}).

\bibitem[{\citenamefont{Behnia et~al.}(2007)\citenamefont{Behnia, Balicas, and
  Kopelevich}}]{Behnia_Bi_Frac_Science07}
\bibinfo{author}{\bibfnamefont{K.}~\bibnamefont{Behnia}},
  \bibinfo{author}{\bibfnamefont{L.}~\bibnamefont{Balicas}}, \bibnamefont{and}
  \bibinfo{author}{\bibfnamefont{Y.}~\bibnamefont{Kopelevich}},
  \bibinfo{journal}{Science} \textbf{\bibinfo{volume}{317}},
  \bibinfo{pages}{1729} (\bibinfo{year}{2007}).

\bibitem[{\citenamefont{Seradjeh
  et~al.}(2009{\natexlab{b}})\citenamefont{Seradjeh, Wu, and
  Phillips}}]{Seradjeh_Bi_HiB_PRL09}
\bibinfo{author}{\bibfnamefont{B.}~\bibnamefont{Seradjeh}},
  \bibinfo{author}{\bibfnamefont{J.}~\bibnamefont{Wu}}, \bibnamefont{and}
  \bibinfo{author}{\bibfnamefont{P.}~\bibnamefont{Phillips}},
  \bibinfo{journal}{Phys.\ Rev.\ Lett.} \textbf{\bibinfo{volume}{103}},
  \bibinfo{pages}{136803} (\bibinfo{year}{2009}{\natexlab{b}}).

\bibitem[{\citenamefont{Sun et~al.}(2009)\citenamefont{Sun, Yao, Fradkin, and
  Kivelson}}]{Sun_TI_SponSymBrk_PRL09}
\bibinfo{author}{\bibfnamefont{K.}~\bibnamefont{Sun}},
  \bibinfo{author}{\bibfnamefont{H.}~\bibnamefont{Yao}},
  \bibinfo{author}{\bibfnamefont{E.}~\bibnamefont{Fradkin}}, \bibnamefont{and}
  \bibinfo{author}{\bibfnamefont{S.~A.} \bibnamefont{Kivelson}},
  \bibinfo{journal}{Phys.\ Rev.\ Lett.} \textbf{\bibinfo{volume}{103}},
  \bibinfo{pages}{046811} (\bibinfo{year}{2009}).

\bibitem[{\citenamefont{Ran et~al.}(to be published)\citenamefont{Ran, Yao, and
  Vishwanath}}]{Ran_TI_Stripes_10}
\bibinfo{author}{\bibfnamefont{Y.}~\bibnamefont{Ran}},
  \bibinfo{author}{\bibfnamefont{H.}~\bibnamefont{Yao}}, \bibnamefont{and}
  \bibinfo{author}{\bibfnamefont{A.}~\bibnamefont{Vishwanath}},
  \bibinfo{journal}{arxiv:1003.0901}  (\bibinfo{year}{to be published}).

\bibitem[{\citenamefont{Raghu et~al.}(2010)\citenamefont{Raghu, Chung, Qi, and
  Zhang}}]{Raghu_HlcLqd_PRL10}
\bibinfo{author}{\bibfnamefont{S.}~\bibnamefont{Raghu}},
  \bibinfo{author}{\bibfnamefont{S.~B.} \bibnamefont{Chung}},
  \bibinfo{author}{\bibfnamefont{X.-L.} \bibnamefont{Qi}}, \bibnamefont{and}
  \bibinfo{author}{\bibfnamefont{S.-C.} \bibnamefont{Zhang}},
  \bibinfo{journal}{Phys.\ Rev.\ Lett.} \textbf{\bibinfo{volume}{104}},
  \bibinfo{pages}{116401} (\bibinfo{year}{2010}).

\bibitem[{\citenamefont{Chen and Raikh}(1999)}]{ChenRaikh_PRB99}
\bibinfo{author}{\bibfnamefont{G.~H.} \bibnamefont{Chen}} \bibnamefont{and}
  \bibinfo{author}{\bibfnamefont{M.~E.} \bibnamefont{Raikh}},
  \bibinfo{journal}{Phys.\ Rev.\ B} \textbf{\bibinfo{volume}{60}},
  \bibinfo{pages}{4826} (\bibinfo{year}{1999}).

\bibitem[{\citenamefont{Glazov and Ivchenko}(2002)}]{Glazov_NATO02}
\bibinfo{author}{\bibfnamefont{M.~M.} \bibnamefont{Glazov}} \bibnamefont{and}
  \bibinfo{author}{\bibfnamefont{E.~L.} \bibnamefont{Ivchenko}},
  \bibinfo{journal}{Proc. NATO Advanced Research Workshop, St.-Petersburg,
  Russia}  (\bibinfo{year}{2002}).

\bibitem[{\citenamefont{Shekhter et~al.}(2005)\citenamefont{Shekhter, Khodas,
  and Finkel'stein}}]{Shekhter_2DEG_ESR_ee_PRB05}
\bibinfo{author}{\bibfnamefont{A.}~\bibnamefont{Shekhter}},
  \bibinfo{author}{\bibfnamefont{M.}~\bibnamefont{Khodas}}, \bibnamefont{and}
  \bibinfo{author}{\bibfnamefont{A.~M.} \bibnamefont{Finkel'stein}},
  \bibinfo{journal}{Phys.\ Rev.\ B} \textbf{\bibinfo{volume}{71}},
  \bibinfo{pages}{165329} (\bibinfo{year}{2005}).

\bibitem[{\citenamefont{Hankiewicz and Vignale}(2006)}]{Hankiewicz_PRB06}
\bibinfo{author}{\bibfnamefont{E.~M.} \bibnamefont{Hankiewicz}}
  \bibnamefont{and} \bibinfo{author}{\bibfnamefont{G.}~\bibnamefont{Vignale}},
  \bibinfo{journal}{Phys.\ Rev.\ B} \textbf{\bibinfo{volume}{73}},
  \bibinfo{pages}{115339} (\bibinfo{year}{2006}).

\bibitem[{\citenamefont{Tse and Sarma}(2007)}]{Tse_SO_Drag_PRB07}
\bibinfo{author}{\bibfnamefont{W.-K.} \bibnamefont{Tse}} \bibnamefont{and}
  \bibinfo{author}{\bibfnamefont{S.~D.} \bibnamefont{Sarma}},
  \bibinfo{journal}{Phys.\ Rev.\ B} \textbf{\bibinfo{volume}{75}},
  \bibinfo{pages}{045333} (\bibinfo{year}{2007}).

\bibitem[{\citenamefont{$\dot{{\rm Z}}$ak et~al.}(2010)\citenamefont{$\dot{{\rm
  Z}}$ak, Maslov, and Loss}}]{Zak_2DEG_SpinSusc_ee_PRB10}
\bibinfo{author}{\bibfnamefont{R.~A.} \bibnamefont{$\dot{{\rm Z}}$ak}},
  \bibinfo{author}{\bibfnamefont{D.~L.} \bibnamefont{Maslov}},
  \bibnamefont{and} \bibinfo{author}{\bibfnamefont{D.}~\bibnamefont{Loss}},
  \bibinfo{journal}{Phys.\ Rev.\ B} \textbf{\bibinfo{volume}{82}},
  \bibinfo{pages}{115415} (\bibinfo{year}{2010}).

\bibitem[{\citenamefont{Culcer}(to appear in Physica E)}]{Culcer_TI_PhysE11}
\bibinfo{author}{\bibfnamefont{D.}~\bibnamefont{Culcer}},
  \bibinfo{journal}{arXiv:1108.3076}  (\bibinfo{year}{to appear in Physica E}).

\bibitem[{\citenamefont{Analytis et~al.}(2010)\citenamefont{Analytis, McDonald,
  Riggs, Chu, Boebinger, and Fisher}}]{Analytis_Bi2Se3_QL_SfcStt_NP10}
\bibinfo{author}{\bibfnamefont{J.~G.} \bibnamefont{Analytis}},
  \bibinfo{author}{\bibfnamefont{R.~D.} \bibnamefont{McDonald}},
  \bibinfo{author}{\bibfnamefont{S.~C.} \bibnamefont{Riggs}},
  \bibinfo{author}{\bibfnamefont{J.-H.} \bibnamefont{Chu}},
  \bibinfo{author}{\bibfnamefont{G.~S.} \bibnamefont{Boebinger}},
  \bibnamefont{and} \bibinfo{author}{\bibfnamefont{I.~R.}
  \bibnamefont{Fisher}}, \bibinfo{journal}{Nat. Phys.}
  \textbf{\bibinfo{volume}{6}}, \bibinfo{pages}{960} (\bibinfo{year}{2010}).

\bibitem[{\citenamefont{Chen et~al.}(2011)\citenamefont{Chen, He, Wu, Ji, Lu,
  Shi, Smet, and Li}}]{Chen_Bi2Se3_Tunable_PRB11}
\bibinfo{author}{\bibfnamefont{J.}~\bibnamefont{Chen}},
  \bibinfo{author}{\bibfnamefont{X.}~\bibnamefont{He}},
  \bibinfo{author}{\bibfnamefont{K.}~\bibnamefont{Wu}},
  \bibinfo{author}{\bibfnamefont{Z.}~\bibnamefont{Ji}},
  \bibinfo{author}{\bibfnamefont{L.}~\bibnamefont{Lu}},
  \bibinfo{author}{\bibfnamefont{J.}~\bibnamefont{Shi}},
  \bibinfo{author}{\bibfnamefont{J.}~\bibnamefont{Smet}}, \bibnamefont{and}
  \bibinfo{author}{\bibfnamefont{Y.}~\bibnamefont{Li}},
  \bibinfo{journal}{Phys.\ Rev.\ B} \textbf{\bibinfo{volume}{83}},
  \bibinfo{pages}{241304} (\bibinfo{year}{2011}).

\bibitem[{\citenamefont{Sacepe et~al.}(to be published)\citenamefont{Sacepe,
  Oostinga, Li, Ubaldini, Couto, Giannini, and Morpurgo}}]{Sacepe_TI_NS_11}
\bibinfo{author}{\bibfnamefont{B.}~\bibnamefont{Sacepe}},
  \bibinfo{author}{\bibfnamefont{J.}~\bibnamefont{Oostinga}},
  \bibinfo{author}{\bibfnamefont{J.}~\bibnamefont{Li}},
  \bibinfo{author}{\bibfnamefont{A.}~\bibnamefont{Ubaldini}},
  \bibinfo{author}{\bibfnamefont{N.}~\bibnamefont{Couto}},
  \bibinfo{author}{\bibfnamefont{E.}~\bibnamefont{Giannini}}, \bibnamefont{and}
  \bibinfo{author}{\bibfnamefont{A.}~\bibnamefont{Morpurgo}},
  \bibinfo{journal}{arXiv:1101.2352}  (\bibinfo{year}{to be published}).

\bibitem[{\citenamefont{Kim et~al.}(to be published)\citenamefont{Kim, Cho,
  Butch, Syers, Kirshenbaum, Paglione, and Fuhrer}}]{Kim_TI_Gate_MinCond_11}
\bibinfo{author}{\bibfnamefont{D.}~\bibnamefont{Kim}},
  \bibinfo{author}{\bibfnamefont{S.}~\bibnamefont{Cho}},
  \bibinfo{author}{\bibfnamefont{N.~P.} \bibnamefont{Butch}},
  \bibinfo{author}{\bibfnamefont{P.}~\bibnamefont{Syers}},
  \bibinfo{author}{\bibfnamefont{K.}~\bibnamefont{Kirshenbaum}},
  \bibinfo{author}{\bibfnamefont{J.}~\bibnamefont{Paglione}}, \bibnamefont{and}
  \bibinfo{author}{\bibfnamefont{M.~S.} \bibnamefont{Fuhrer}},
  \bibinfo{journal}{arXiv:1105.1410}  (\bibinfo{year}{to be published}).

\bibitem[{\citenamefont{B\"uttner et~al.}(2011)\citenamefont{B\"uttner, Liu,
  Tkachov, Novik, Br\"une, Buhmann, Hankiewicz, Recher, Trauzettel, Zhang
  et~al.}}]{Buettner_HgTe_ZeroGapQW_NP11}
\bibinfo{author}{\bibfnamefont{B.}~\bibnamefont{B\"uttner}},
  \bibinfo{author}{\bibfnamefont{C.}~\bibnamefont{Liu}},
  \bibinfo{author}{\bibfnamefont{G.}~\bibnamefont{Tkachov}},
  \bibinfo{author}{\bibfnamefont{E.}~\bibnamefont{Novik}},
  \bibinfo{author}{\bibfnamefont{C.}~\bibnamefont{Br\"une}},
  \bibinfo{author}{\bibfnamefont{H.}~\bibnamefont{Buhmann}},
  \bibinfo{author}{\bibfnamefont{E.~M.} \bibnamefont{Hankiewicz}},
  \bibinfo{author}{\bibfnamefont{P.}~\bibnamefont{Recher}},
  \bibinfo{author}{\bibfnamefont{B.}~\bibnamefont{Trauzettel}},
  \bibinfo{author}{\bibfnamefont{S.}~\bibnamefont{Zhang}},
  \bibnamefont{et~al.}, \bibinfo{journal}{Nat.\ Phys.}
  \textbf{\bibinfo{volume}{7}}, \bibinfo{pages}{418} (\bibinfo{year}{2011}).

\bibitem[{\citenamefont{Br\"une et~al.}(2011)\citenamefont{Br\"une, Liu, Novik,
  Hankiewicz, Buhmann, Chen, Qi, Shen, Zhang, and
  Molenkamp}}]{Bruene_StrainedHgTe_QHE_PRL11}
\bibinfo{author}{\bibfnamefont{C.}~\bibnamefont{Br\"une}},
  \bibinfo{author}{\bibfnamefont{C.}~\bibnamefont{Liu}},
  \bibinfo{author}{\bibfnamefont{E.}~\bibnamefont{Novik}},
  \bibinfo{author}{\bibfnamefont{E.}~\bibnamefont{Hankiewicz}},
  \bibinfo{author}{\bibfnamefont{H.}~\bibnamefont{Buhmann}},
  \bibinfo{author}{\bibfnamefont{Y.~L.} \bibnamefont{Chen}},
  \bibinfo{author}{\bibfnamefont{X.~L.} \bibnamefont{Qi}},
  \bibinfo{author}{\bibfnamefont{Z.}~\bibnamefont{Shen}},
  \bibinfo{author}{\bibfnamefont{S.}~\bibnamefont{Zhang}}, \bibnamefont{and}
  \bibinfo{author}{\bibfnamefont{L.}~\bibnamefont{Molenkamp}},
  \bibinfo{journal}{Phys.\ Rev.\ Lett.} \textbf{\bibinfo{volume}{106}},
  \bibinfo{pages}{126803} (\bibinfo{year}{2011}).

\bibitem[{\citenamefont{Culcer et~al.}(2010)\citenamefont{Culcer, Hwang,
  Stanescu, and {Das Sarma}}}]{Culcer_TI_Kineq_PRB10}
\bibinfo{author}{\bibfnamefont{D.}~\bibnamefont{Culcer}},
  \bibinfo{author}{\bibfnamefont{E.~H.} \bibnamefont{Hwang}},
  \bibinfo{author}{\bibfnamefont{T.~D.} \bibnamefont{Stanescu}},
  \bibnamefont{and} \bibinfo{author}{\bibfnamefont{S.}~\bibnamefont{{Das
  Sarma}}}, \bibinfo{journal}{Phys.\ Rev.\ B} \textbf{\bibinfo{volume}{82}},
  \bibinfo{pages}{155457} (\bibinfo{year}{2010}).

\bibitem[{\citenamefont{Zyuzin et~al.}(2011)\citenamefont{Zyuzin, Hook, and
  Burkov}}]{Zyuzin_TI_B_parallel_QPT_PRB11}
\bibinfo{author}{\bibfnamefont{A.}~\bibnamefont{Zyuzin}},
  \bibinfo{author}{\bibfnamefont{M.}~\bibnamefont{Hook}}, \bibnamefont{and}
  \bibinfo{author}{\bibfnamefont{A.}~\bibnamefont{Burkov}},
  \bibinfo{journal}{Phys.\ Rev.\ B} \textbf{\bibinfo{volume}{83}},
  \bibinfo{pages}{245428} (\bibinfo{year}{2011}).

\bibitem[{\citenamefont{Culcer and Winkler}(2008)}]{Culcer_Gfn_Transp_PRB08}
\bibinfo{author}{\bibfnamefont{D.}~\bibnamefont{Culcer}} \bibnamefont{and}
  \bibinfo{author}{\bibfnamefont{R.}~\bibnamefont{Winkler}},
  \bibinfo{journal}{Phys.\ Rev.\ B} \textbf{\bibinfo{volume}{78}},
  \bibinfo{pages}{235417} (\bibinfo{year}{2008}).

\bibitem[{\citenamefont{Culcer and Winkler}(2009)}]{Culcer_Bil_PRB09}
\bibinfo{author}{\bibfnamefont{D.}~\bibnamefont{Culcer}} \bibnamefont{and}
  \bibinfo{author}{\bibfnamefont{R.}~\bibnamefont{Winkler}},
  \bibinfo{journal}{Phys.\ Rev.\ B} \textbf{\bibinfo{volume}{79}},
  \bibinfo{pages}{165422} (\bibinfo{year}{2009}).

\bibitem[{\citenamefont{D'Amico and Vignale}(2000)}]{Amico_PRB00}
\bibinfo{author}{\bibfnamefont{I.}~\bibnamefont{D'Amico}} \bibnamefont{and}
  \bibinfo{author}{\bibfnamefont{G.}~\bibnamefont{Vignale}},
  \bibinfo{journal}{Phys.\ Rev.\ B} \textbf{\bibinfo{volume}{62}},
  \bibinfo{pages}{4853} (\bibinfo{year}{2000}).

\bibitem[{\citenamefont{{Das Sarma} et~al.}(2011)\citenamefont{{Das Sarma},
  Adam, Hwang, and Rossi}}]{SDS_Gfn_RMP11}
\bibinfo{author}{\bibfnamefont{S.}~\bibnamefont{{Das Sarma}}},
  \bibinfo{author}{\bibfnamefont{S.}~\bibnamefont{Adam}},
  \bibinfo{author}{\bibfnamefont{E.~H.} \bibnamefont{Hwang}}, \bibnamefont{and}
  \bibinfo{author}{\bibfnamefont{E.}~\bibnamefont{Rossi}},
  \bibinfo{journal}{Rev.\ Mod.\ Phys.} \textbf{\bibinfo{volume}{83}},
  \bibinfo{pages}{407} (\bibinfo{year}{2011}).

\bibitem[{\citenamefont{Hwang and {Das
  Sarma}}(2007)}]{Hwang_Gfn_Screening_PRB07}
\bibinfo{author}{\bibfnamefont{E.~H.} \bibnamefont{Hwang}} \bibnamefont{and}
  \bibinfo{author}{\bibfnamefont{S.}~\bibnamefont{{Das Sarma}}},
  \bibinfo{journal}{Phys.\ Rev.\ B} \textbf{\bibinfo{volume}{75}},
  \bibinfo{pages}{205418} (\bibinfo{year}{2007}).

\bibitem[{\citenamefont{Vasko and Raichev}(2005)}]{Vasko}
\bibinfo{author}{\bibfnamefont{F.~T.} \bibnamefont{Vasko}} \bibnamefont{and}
  \bibinfo{author}{\bibfnamefont{O.~E.} \bibnamefont{Raichev}},
  \emph{\bibinfo{title}{Quantum Kinetic Theory and Applications}}
  (\bibinfo{publisher}{Springer}, \bibinfo{address}{New York},
  \bibinfo{year}{2005}).

\bibitem[{\citenamefont{Tse and MacDonald}(2009)}]{Tse_Bil_OptCond_PRB09}
\bibinfo{author}{\bibfnamefont{W.-K.} \bibnamefont{Tse}} \bibnamefont{and}
  \bibinfo{author}{\bibfnamefont{A.~H.} \bibnamefont{MacDonald}},
  \bibinfo{journal}{Phys.\ Rev.\ B} \textbf{\bibinfo{volume}{80}},
  \bibinfo{pages}{195418} (\bibinfo{year}{2009}).

\bibitem[{\citenamefont{Teo et~al.}(2008)\citenamefont{Teo, Fu, and
  Kane}}]{Teo_BiSb_SfcStt_PRB08}
\bibinfo{author}{\bibfnamefont{J.~C.} \bibnamefont{Teo}},
  \bibinfo{author}{\bibfnamefont{L.}~\bibnamefont{Fu}}, \bibnamefont{and}
  \bibinfo{author}{\bibfnamefont{C.}~\bibnamefont{Kane}},
  \bibinfo{journal}{Phys.\ Rev.\ B} \textbf{\bibinfo{volume}{78}},
  \bibinfo{pages}{045426} (\bibinfo{year}{2008}).

\bibitem[{\citenamefont{Zhang et~al.}(2009{\natexlab{a}})\citenamefont{Zhang,
  Liu, Qi, Deng, Dai, Zhang, and Fang}}]{Zhang_BiSb_SfcStt_PRB09}
\bibinfo{author}{\bibfnamefont{H.-J.} \bibnamefont{Zhang}},
  \bibinfo{author}{\bibfnamefont{C.-X.} \bibnamefont{Liu}},
  \bibinfo{author}{\bibfnamefont{X.-L.} \bibnamefont{Qi}},
  \bibinfo{author}{\bibfnamefont{X.-Y.} \bibnamefont{Deng}},
  \bibinfo{author}{\bibfnamefont{X.}~\bibnamefont{Dai}},
  \bibinfo{author}{\bibfnamefont{S.-C.} \bibnamefont{Zhang}}, \bibnamefont{and}
  \bibinfo{author}{\bibfnamefont{Z.}~\bibnamefont{Fang}},
  \bibinfo{journal}{Phys.\ Rev.\ B} \textbf{\bibinfo{volume}{80}},
  \bibinfo{pages}{085307} (\bibinfo{year}{2009}{\natexlab{a}}).

\bibitem[{\citenamefont{Zhang et~al.}(2009{\natexlab{b}})\citenamefont{Zhang,
  Liu, Qi, Dai, Fang, and Zhang}}]{Zhang_TI_BandStr_NP09}
\bibinfo{author}{\bibfnamefont{H.}~\bibnamefont{Zhang}},
  \bibinfo{author}{\bibfnamefont{C.-X.} \bibnamefont{Liu}},
  \bibinfo{author}{\bibfnamefont{X.-L.} \bibnamefont{Qi}},
  \bibinfo{author}{\bibfnamefont{X.}~\bibnamefont{Dai}},
  \bibinfo{author}{\bibfnamefont{Z.}~\bibnamefont{Fang}}, \bibnamefont{and}
  \bibinfo{author}{\bibfnamefont{S.-C.} \bibnamefont{Zhang}},
  \bibinfo{journal}{Nat.\ Phys.} \textbf{\bibinfo{volume}{5}},
  \bibinfo{pages}{438} (\bibinfo{year}{2009}{\natexlab{b}}).

\bibitem[{\citenamefont{Zhang et~al.}(2010)\citenamefont{Zhang, Yu, Zhang, Dai,
  and Fang}}]{Zhang_TI_1stPrinc_NJP10}
\bibinfo{author}{\bibfnamefont{W.}~\bibnamefont{Zhang}},
  \bibinfo{author}{\bibfnamefont{R.}~\bibnamefont{Yu}},
  \bibinfo{author}{\bibfnamefont{H.-J.} \bibnamefont{Zhang}},
  \bibinfo{author}{\bibfnamefont{X.}~\bibnamefont{Dai}}, \bibnamefont{and}
  \bibinfo{author}{\bibfnamefont{Z.}~\bibnamefont{Fang}}, \bibinfo{journal}{New
  J. Phys.} \textbf{\bibinfo{volume}{12}}, \bibinfo{pages}{065013}
  (\bibinfo{year}{2010}).

\bibitem[{\citenamefont{Xiao et~al.}(2010)\citenamefont{Xiao, Yao, Feng, Wen,
  Zhu, Chen, Stocks, and Zhang}}]{Xiao_Heusler_PRL10}
\bibinfo{author}{\bibfnamefont{D.}~\bibnamefont{Xiao}},
  \bibinfo{author}{\bibfnamefont{Y.}~\bibnamefont{Yao}},
  \bibinfo{author}{\bibfnamefont{W.}~\bibnamefont{Feng}},
  \bibinfo{author}{\bibfnamefont{J.}~\bibnamefont{Wen}},
  \bibinfo{author}{\bibfnamefont{W.}~\bibnamefont{Zhu}},
  \bibinfo{author}{\bibfnamefont{X.-Q.} \bibnamefont{Chen}},
  \bibinfo{author}{\bibfnamefont{G.~M.} \bibnamefont{Stocks}},
  \bibnamefont{and} \bibinfo{author}{\bibfnamefont{Z.}~\bibnamefont{Zhang}},
  \bibinfo{journal}{Phys.\ Rev.\ Lett.} \textbf{\bibinfo{volume}{105}},
  \bibinfo{pages}{096404} (\bibinfo{year}{2010}).

\bibitem[{\citenamefont{Feng et~al.}(2011)\citenamefont{Feng, Xiao, Ding, and
  Yao}}]{Feng_TI_Chalcop_PRL11}
\bibinfo{author}{\bibfnamefont{W.}~\bibnamefont{Feng}},
  \bibinfo{author}{\bibfnamefont{D.}~\bibnamefont{Xiao}},
  \bibinfo{author}{\bibfnamefont{J.}~\bibnamefont{Ding}}, \bibnamefont{and}
  \bibinfo{author}{\bibfnamefont{Y.}~\bibnamefont{Yao}},
  \bibinfo{journal}{Phys.\ Rev.\ Lett.} \textbf{\bibinfo{volume}{106}},
  \bibinfo{pages}{016402} (\bibinfo{year}{2011}).

\end{thebibliography}

\end{document}